\documentclass[a4paper,11pt]{article}
\usepackage{pos}
\usepackage{units}
\usepackage{feynmp}
\usepackage{amsmath,bbm}
\usepackage{slashed}
\usepackage{graphicx}
\usepackage{caption}
\usepackage{subcaption}

\renewcommand{\to}{\rightarrow}

\newcommand{\Lag}{\mathscr{L}}
\newcommand{\de}{\partial}

\renewcommand{\O}{\mathbf{O}}

\renewcommand{\c}{\mathbf{c}}

\newcommand{\beq}{\begin{equation}}
\newcommand{\eeq}{\end{equation}}
\newcommand{\bea}{\begin{eqnarray}}
\newcommand{\eea}{\end{eqnarray}}
\renewcommand{\[}{\begin{equation}}
\renewcommand{\]}{\end{equation}}

\title{ One-loop corrections to ALP couplings}
\author*[a]{ J.~Bonilla}
\author[b]{I.~Brivio}
\author[a]{M.B.~Gavela} 
\author[c,d]{V.~Sanz}

\affiliation[a]{Departamento de F\'isica Te\'orica, Universidad Aut\'onoma de Madrid, and Instituto de F\'isica Te\'orica IFT-UAM/CSIC, 
Cantoblanco, E-28049, Madrid, Spain}
\affiliation[b]{Institut f\"ur Theoretische Physik, Universit\"at Heidelberg, Philosophenweg 16,\\ D-69120 Heidelberg, Germany}
\affiliation[c]{Instituto de F{\' i}sica Corpuscular (IFIC), Universidad de Valencia-CSIC, E-46980 Valencia, Spain}
\affiliation[d]{Department of Physics and Astronomy, University of Sussex, Brighton BN1 9QH, UK}
\emailAdd{jesus.bonilla@uam.es}
\emailAdd{brivio@thphys.uni-heidelberg.de}
\emailAdd{belen.gavela@uam.es}
\emailAdd{veronica.sanz@uv.es}

\abstract{

We derive the one-loop contributions to ALP-SM effective couplings, 
including all finite corrections. 
The  complete leading-order --dimension five-- effective linear Lagrangian is 
considered.  
The ALP is left off-shell, which is of particular impact on LHC and accelerator 
searches of 
ALP couplings to 
$\gamma\gamma$, $ZZ$, $Z\gamma$, $WW$, gluons and fermions. All  results are obtained 
in the covariant $R_\xi$ gauge.
A few phenomenological consequences are also explored as illustration,  with flavour diagonal channels in the case of fermions: in particular, we explore constraints on the coupling of the ALP to top quarks, that can be extracted from LHC data, from astrophysical  sources and from Dark Matter direct detection experiments  such as PandaX, LUX and XENON1T.
}

\FullConference{%
  *** The European Physical Society Conference on High Energy Physics (EPS-HEP2021), ***\\
  *** 26-30 July 2021 ***\\
  *** Online conference, jointly organized by Universität Hamburg and the research center DESY ***
}

\begin{document}
\maketitle

\paragraph{Introduction}Here we summarize the work presented in Ref.~\cite{Bonilla:2021ufe}, in which we explore at one-loop order all possible  CP-even operators coupling one pseudoscalar ALP to SM fields at next-to leading order of the linear effective field theory formulation, i.e. mass dimension five operators.

We provide the complete one-loop corrections, i.e.  divergent and finite contributions,  
for an off-shell  ALP and on-shell SM fields. All our computations are performed in the covariant $R_\xi$ gauge.   
The only restriction on fermions is that only flavour diagonal channels are computed. CKM quark mixing and neutrino masses are disregarded.   

The results are timely because the level of experimental sensitivity to several ALP-SM couplings has reached a level where one-loop corrections are necessary, and in some cases they are the best tool to constraint some couplings.

\paragraph{Effective Lagrangian}A complete  and non-redundant ALP effective Lagrangian is given at $\mathcal{O} (1/f_a)$  by $\Lag_{ALP} = \Lag_{SM} + \Lag_a^{\rm total}$, where $\Lag_{SM}$ denotes the SM Lagrangian. The most general CP-conserving ALP effective Lagrangian  $\Lag_a^\text{total}$, including {\it bosonic and fermionic} ALP couplings~\cite{Georgi:1986df, Choi:1986zw}, admits many possible choices of basis.  A complete is that defined  by the Lagrangian
\begin{equation}
 \Lag_a^\text{total}\,= \frac12 \de_\mu a\de^\mu a + \frac{m_a^2}{2}a^2+\,c_{\tilde{W}}\O_{\tilde{W}}+c_{\tilde{B}}\O_{\tilde{B}}+c_{\tilde{G}}\O_{\tilde{G}}+ \sum_{\text{f}=u,\,d,\,e, \, Q, \, L}\,\text{\bf{c}}_{\text{f}}\, \O_{\text{f}}\,,
\label{general-NLOLag-lin}
\end{equation}
where $\O_{\tilde{X}} = - a/f_a \, X \tilde{X}$ and $\O_\text{f}^{ij} = {\de_\mu a}/{f_a} \, \big(\bar{\text{f}}^i \gamma_\mu  \text{f}^j \big)$. The three coefficients  $c_{\tilde X}$ of the gauge-anomalous operators   are real scalar quantities, while $\c_{\text{f}}$ are 3x3 hermitian matrices in flavour space. In addition, because of the assumption of CP conservation, they obey $\c_{\text{f }}= \c_{\text{f }}^T$. Moreover, in our calculation we neglect flavor mixing (assuming CKM $=\mathbbm{1}$). In this limit, the 6 flavor-diagonal degrees of freedom in the left-handed fermion operators are redundant and are therefore excluded. If CKM mixing is restored, 2 of the diagonal elements for left-handed quarks need to be reintroduced.

\begin{figure}[b]
\centering
\includegraphics[width=\textwidth]{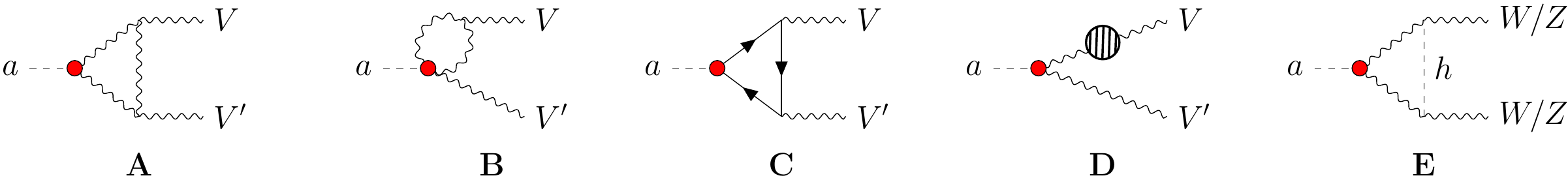}
\caption{One-loop diagrams contributing to bosonic ALP interactions at one-loop (the corresponding diagrams with Goldstone bosons and the diagrams exchanging the gauge boson legs are left implicit), where $V$ and $V'$ are either a  gluon,  a photon,
a $Z$ boson or a $W$ boson. Diagram \textbf{D} includes all SM corrections to Gauge boson external legs. The last diagram only corrects the couplings with $Z$ and $W$ bosons.}
\label{correctionsgammaZ}
\end{figure}

\begin{figure}[h]
\centering
\includegraphics[width=\textwidth]{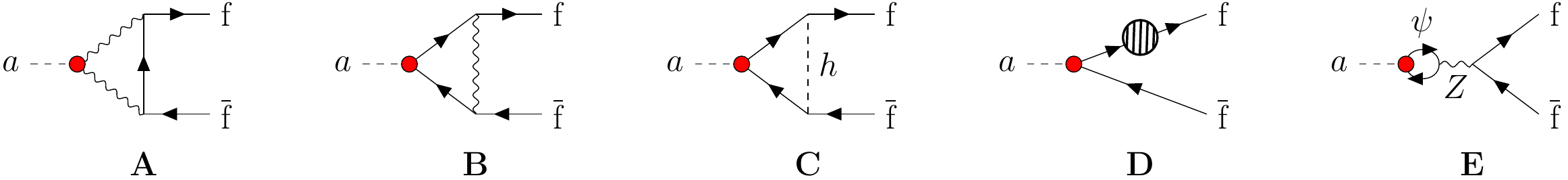}
\caption{One-loop diagrams contributing to $\c_{\text{f}}$  at one-loop (plus the corresponding diagrams with Goldstone bosons). The wavy lines denote gauge bosons: gluons, photons, $W$ and $Z$ bosons. Diagram \textbf{D} includes all SM corrections to fermion external legs. In the last diagram all SM fermions participate in the loop.}
\label{correctionsferm}
\end{figure}

\paragraph{Complete one-loop contributions to ALP couplings}We compute the one-loop contributions to the phenomenological ALP couplings,  including all finite corrections. 1-loop diagrams contributing to bosonic and fermionic couplings are represented in Fig.~\ref{correctionsgammaZ} and Fig.~\ref{correctionsferm}, respectively.

The ALP field will be left off-shell (which is of practical interest for collider and other searches away from the ALP resonance), while the external SM fields will be considered on-shell. For channels with external fermions, we only provide corrections to the flavour diagonal ones. CKM quark mixing's disregarded in the loop corrections to all couplings. All computations have been carried out in the covariant $R_{\xi}$-gauge, with the help of  {\tt Mathematica} packages {\tt FeynCalc} and {\tt Package-X}~\cite{Shtabovenko:2020gxv,Patel:2016fam}. 

The complete analytic results are lengthy and are not shown here. They are provided in \href{https://notebookarchive.org/2021-07-9otlr9o}{NotebookArchive} in addition with some useful intermediate steps. In the next section we present an example on their use to set new bounds on the ALP parameter space, in particular, to the ALP-top quark interaction.

 \paragraph{Some phenomenological consequences: bounds on the ALP-top quark coupling}We consider two examples of situations in which experimental test are able to probe loop corrections: high-energy gluon-initiated production of ALPs at LHC, and very precise low-energy searches for ALPs which rely on couplings to electrons. In both cases we assume that the ALP only couples to top quarks at tree level, so that the ALP-gluon and ALP-electron interactions emerge at one-loop by top-quark mediated processes depicted in diagram \textbf{C} in Fig.~\ref{correctionsgammaZ} and diagram \textbf{E} in Fig.~\ref{correctionsferm} respectively.

\begin{figure}[b]
\begin{subfigure}{.47\textwidth}
  \includegraphics[width=6.6cm]{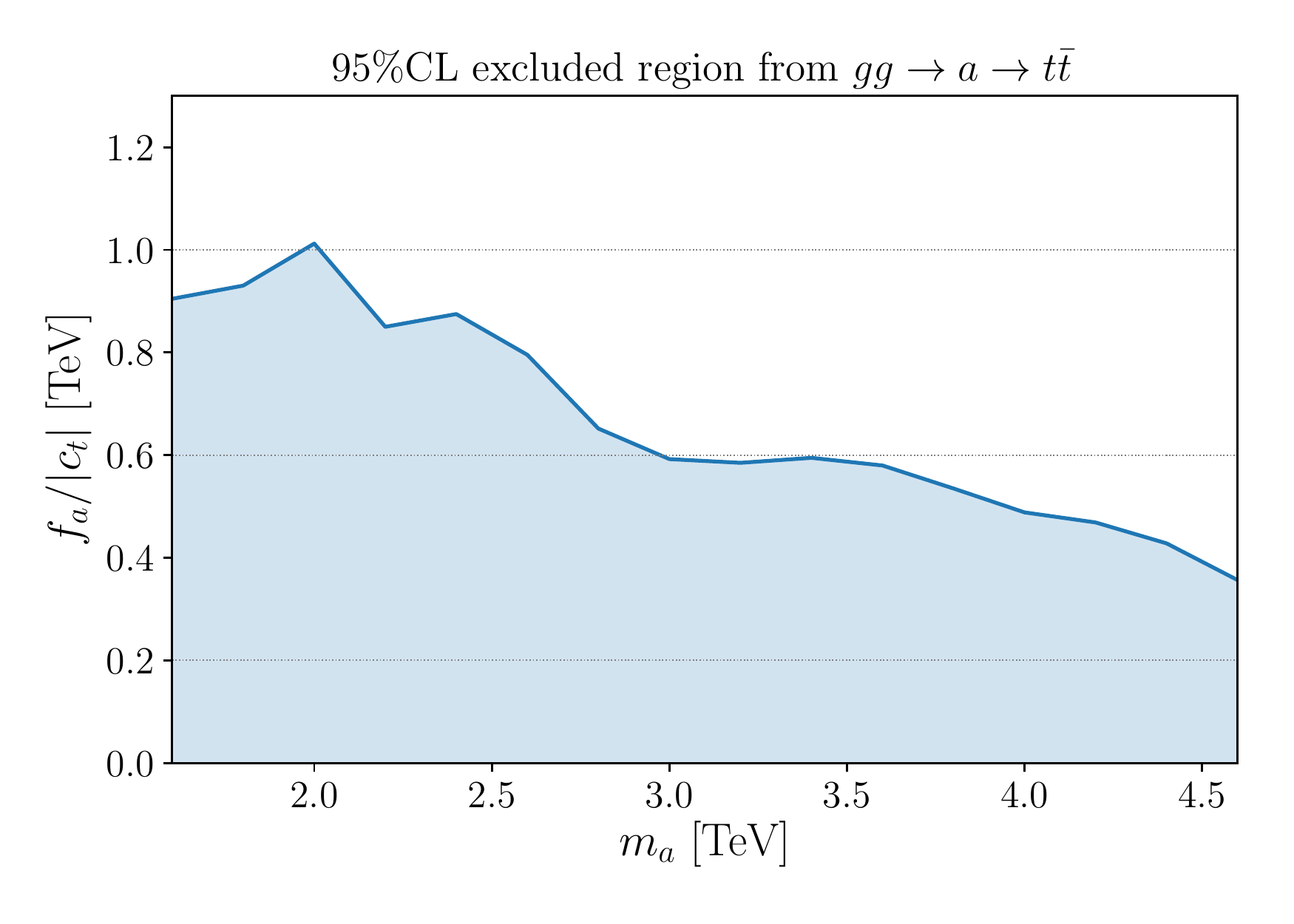}
  \caption{Limits on $f_a/|c_t|$, as a function of the ALP mass, \\
  extracted from the all-hadronic $t\bar t$ resonance search \\
  by ATLAS of  Ref.~\cite{ATLAS:2020lks}.}
  \label{fatoplimit}
\end{subfigure}%
\begin{subfigure}{.5\textwidth}
  \includegraphics[width=8.8cm]{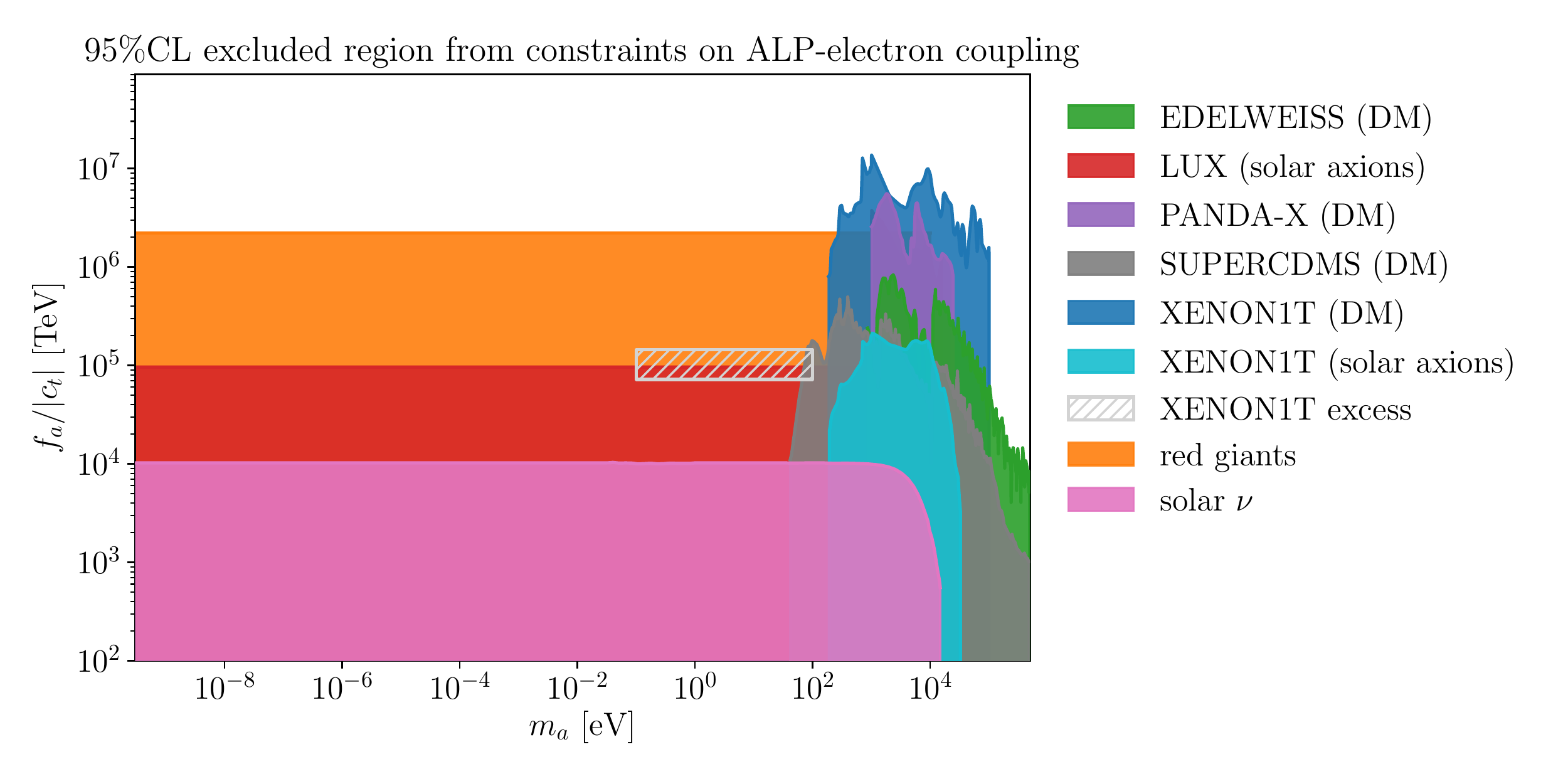}
  \caption{Limits on $f_a/|c_t|$ as a function of the ALP mass, extracted rescaling existing constraints on the ALP-electron coupling, taken from Ref.~\cite{ciaran_o_hare_2020_3932430}.}
  \label{fig:bounds_ee}
\end{subfigure}
\caption{Limits on $f_a/|c_t|$.}
\label{fig:ct}
\end{figure}

\subparagraph{$\bullet$ Gluon-initiated production of ALPs}We examine the resonant gluon-initiated production of an ALP, which later decays in pair of top quarks: $gg\to a\to t\bar t$. For definiteness, here we consider ALPs with $m_a>2m_t$, such that the top-antitop pair can be resonant. This allows us to derive constraints from existing searches for resonances in $t\bar t$ final states, that are at a very mature stage in the LHC collaborations. As an illustration, we re-interpret the recent ATLAS analysis~\cite{ATLAS:2020lks} to set bounds on  $c_t/f_a$, where we define $
c_t \equiv (\c_u)_{33} $.

We have performed a simulation of this process in {\tt MadGraph5\_aMC@NLO}~\cite{Alwall:2011uj}, using an in-house UFO implementation of the Lagrangian in Eq.~\eqref{general-NLOLag-lin}. The distribution obtained (summing signal and interference components) is compared to the difference between measured and predicted number of events in the $m_{t\bar t}$ spectra reported in Ref.~\cite{ATLAS:2020lks}, that is available on~\href{https://www.hepdata.net/record/ins1795076?version=1}{HEPdata}. The results of this re-interpretation are shown in  Fig.~\ref{fatoplimit}.

\subparagraph{$\bullet$ Low energy searches of ALPs}We consider the current limits on the axion-electron coupling collected in Ref.~\cite{ciaran_o_hare_2020_3932430}, that include results from solar axions searches, ALP DM searches as well as astrophysical bounds. At one-loop, this interaction is given by (see also Ref.~\cite{Feng:1997tn}):
\begin{equation} \label{cefromctop}
 c_{e}^{\text{eff}} \simeq  2.48 \, c_t \,  \alpha_{em} \,  \log \left( \frac{\Lambda^2}{m_t^2}  \right) \,,
\end{equation}
where we take $\Lambda=\unit[10^6]{TeV}$   in this equation, to extract the bounds on $f_a/c_t$ shown in Fig.~\ref{fig:bounds_ee}.

\paragraph{Conclusions}Here we have summarized the work in Ref.~\cite{Bonilla:2021ufe}, where we computed  the complete one-loop corrections --thus including all divergent and finite terms-- to all possible couplings in the CP-even base for the $d=5$ ALP linear effective Lagrangian, for a generic off-shell ALP and on-shell SM particles.  These results are publicly available at 
\href{https://notebookarchive.org/2021-07-9otlr9o}{NotebookArchive}.

For instance, we explored how, for heavy ALPs, the ALP-top coupling can be constrained by LHC measurements of top-pair final states, processes which are induced at one-loop by this coupling. We also explored constraints on ALP-top interactions for light ALPs. In this case, the strictest limits are those derived from bounds on the ALP-electron coupling, extracted from astrophysical constraints and from DM Direct Detection searches.

\bibliographystyle{JHEP}
\bibliography{bibliography}

\end{document}